\documentclass[conference]{IEEEtran}
\IEEEoverridecommandlockouts
\pdfoutput=1

\usepackage{tikz}
\usepackage{textcomp}
\usepackage{hyperref}
\usepackage{lipsum}

\usepackage{cite}
\usepackage{amsmath,amssymb,amsfonts}
\usepackage{algorithmic}
\usepackage{algorithm}
\usepackage{graphicx}
\usepackage{textcomp}
\usepackage{xcolor}
\usepackage{multirow}
\usepackage{booktabs}
\usepackage{tabularx}
\def\BibTeX{{\rm B\kern-.05em{\sc i\kern-.025em b}\kern-.08em
    T\kern-.1667em\lower.7ex\hbox{E}\kern-.125emX}}
\begin{document}

\title{Learning to Detect: A Data-driven Approach for Network Intrusion Detection

}

\author{
\IEEEauthorblockN{Zachary Tauscher, Yushan Jiang, Kai Zhang, Jian Wang, Houbing Song}

\IEEEauthorblockA{\textit{Department of Electrical Engineering \& Computer Science} \\
\textit{Embry-Riddle Aeronautical University}\\
Daytona Beach, FL 32114 USA \\
\{tauschez, jiangy2, zhangk3, wangj14\}@my.erau.edu, h.song@ieee.org}





}

\maketitle
\makeatletter
\def\ps@IEEEtitlepagestyle{%
  \def\@oddfoot{\mycopyrightnotice}%
  \def\@oddhead{\hbox{}\@IEEEheaderstyle\leftmark\hfil\thepage}\relax
  \def\@evenhead{\@IEEEheaderstyle\thepage\hfil\leftmark\hbox{}}\relax
  \def\@evenfoot{}%
}
\def\mycopyrightnotice{%
  \begin{minipage}{\textwidth}
  \centering \scriptsize
  Copyright~\copyright~2021 IEEE. Personal use of this material is permitted. Permission from IEEE must be obtained for all other uses, in any current or future media, including\\reprinting/republishing this material for advertising or promotional purposes, creating new collective works, for resale or redistribution to servers or lists, or reuse of any copyrighted component of this work in other works by sending a request to pubs-permissions@ieee.org.
  \end{minipage}
}
\makeatother

\begin{abstract}
With massive data being generated daily and the ever-increasing interconnectivity of the world's Internet infrastructures, a machine learning based intrusion detection system (IDS) has become a vital component to protect our economic and national security. In this paper, we perform a comprehensive study on NSL-KDD, a network traffic dataset, by visualizing patterns and employing different learning-based models to detect cyber attacks. Unlike previous shallow learning and deep learning models that use the single learning model approach for intrusion detection, we adopt a hierarchy strategy, in which the intrusion and normal behavior are classified firstly, and then the specific types of attacks are classified. We demonstrate the advantage of the unsupervised representation learning model in binary intrusion detection tasks. Besides, we alleviate the data imbalance problem with SVM-SMOTE oversampling technique in 4-class classification and further demonstrate the effectiveness and the drawback of the oversampling mechanism with a deep neural network as a base model.
\end{abstract}

\begin{IEEEkeywords}
Intrusion Detection System, Machine Learning, Data Analytics, Computer Networks, NSL-KDD  
\end{IEEEkeywords}

\section{Introduction}
Global communication and networking are commonplace in the current era. Everything from cell phones to thermostats is connected to the internet. A large number of users and devices connected to the internet makes the security risk to these networks only that much greater. The ability to detect and prevent network attacks is vital to maintain the confidentiality, integrity, and availability of our information and communication systems. Network intrusion detection and prevention systems (IDS/IPS) are a critical part of any network or system architecture designed to record and analyze connection behavior to identify possible attacks and report such information to an administrator or prevent the attack entirely.

IDPS technologies vary in their methodologies for detecting intrusions but tend to fall into two specific categories, signature-based and anomaly-based detection. Signature-based IDS systems, also known as misuse IDS, have been the most widely used due to their simplicity and reliability. These types of systems utilize pattern recognition to compare signatures of well-known attacks to current connections \cite{6759203}. Anomaly-based IDS technology analyzes normal network traffic to develop models of normal behavior. Any connections that then deviate from these models are flagged as an intrusion. Anomaly-based IDS often produce a high volume of false positives as any activity that deviates from the normal is flagged. So while signature-based IDS is more often used, anomaly-based IDS has greater potential power, especially as machine learning and AI models continue to develop and become a greater focus in cybersecurity \cite{key:article}.


With the development of capable AI-driven IDS/IPS technologies, there have been various studies investigating and developing data-driven methods. Besides the data analytics and pattern presentations using traditional visualization techniques \cite{turcotte2019unified} and unsupervised K-means clustering \cite{zong2019dimensionality} , the method toward detecting attacks can be mainly divided into two parts, the classical machine learning classifiers and deep learning models. In classical learning methods, several classifiers are utilized and modified for binary and multi-class intrusion detection tasks \cite{naivebayes,negandhi2019intrusionrf,divekar2018benchmarking,hu2013online}, including basic tree methods, Multi-layer Perceptron, and Support Vector Machine, Naive Bayes, Random Forest, and a sophisticated variant of boost-based classifiers. Besides, feature selection is leveraged to choose the informative subset of features to facilitate the performance of classifiers, which is based on different techniques including Flexible Neural Tree \cite{chen2005feature}, visualization techniques of distribution histograms, scatter plots, and information gain \cite{staudemeyer2014extracting}. In deep learning methods, besides the deep neural networks \cite{dnn}, Convolutional Neural Networks \cite{wu2019lunet}, Recurrent Neural Networks \cite{althubiti2018lstm}, and their integration \cite{cnnlstm} are applied to capture certain spatial characteristics and temporal dependencies in an individual or joint manner, for downstream intrusion classification task. Moreover, to tickle the data insufficient issues in some datasets, transfer learning \cite{wu2019transfer} and Variational Autoencoder \cite{yang2020network} are also considered in terms of representation transferring and learning, which generalizes the learning-based methods to a wider range of applications.

\begin{table*}[htbp]
\caption{Attack Category and Its Statistics in NSL-KDD dataset}
\begin{center}
\begin{tabularx}{0.95\textwidth}{lXX}
 \toprule
 \textbf{Attack Category} & 
 \textbf{Description} & \textbf{Attack Type \& Count}\\
 \midrule
 Probe & A process of probing target computer network to find weaknesses in its defense & satan (3633), portsweep (2931), nmap (1493), jpsweep (3599) \\ 
 \midrule
 R2L & Unauthorized access from a remote machine to a local machine & spy (2), phf (4), multihop (7), imap (11), guess\_passwd (53), ftp\_write (8), warezmaster (20), warezclient (890) \\
 \midrule
 U2R & Unauthorized access to local superuser privileges by a local unprivileged user & rootkit (10), perl (3), loadmodule (9), buffer\_overflow (30), \\
 \midrule
 DoS & Oversaturating connection bandwidth or depleting the target's system resources & teardrop (892), smurf (2646), pod (201), neptune (41214), land (18), back (956)\\
 \bottomrule
\end{tabularx}
\label{tab:attack_c}
\end{center}
\end{table*}

\begin{figure*}[htbp]
    \centering
    \includegraphics[scale=.3]{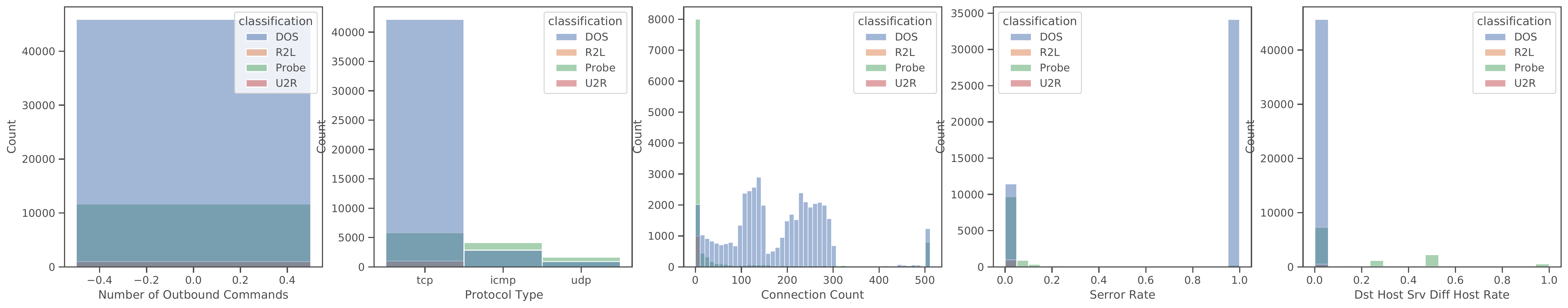}
    \caption{Distribution histograms of specified features in the NSL-KDD data set.}
    \label{fig:Dis}
\end{figure*}

In this paper, we aim at developing a data-driven intrusion detection framework to analyze and classify the patterns of normal network status and various malicious attacks. To be specific, we perform exploration on the NSL-KDD dataset representing real-world network traffic, by visualizing and analyzing potential patterns so that preliminary decision making and manipulation can be taken. Moreover, we adopt a two-stage hierarchy strategy based on machine learning models for intrusion detection tasks, where the attacks with abnormal patterns are first detected from normal samples, then further classified into different types. At the first stage, we adopt supervised classifiers and a representation learning model to detect anomalies. As the intrusion data suffers from a severe imbalance problem, we first leverage an oversampling technique at the second stage, then utilize a deep neural network to classify the attack type in a supervised manner.

\section{Data Exploration}

In this section, we present the exploration of the NSL-KDD dataset, which includes the description, pattern visualization, and analytic. 

\subsection{Dataset Description}
NSL-KDD is the refined version of the KDD'99 \cite{tavallaee2009detailed} data set to solve its inherent problems. For example, it does not contain redundant records so that the model training and evaluation is not biased by high-frequently duplicated records. Moreover, the number of selected records from each difficulty level group is inversely proportional to the percentage of records in the KDD'99 data set, which results in varying classification rates of different machine learning and it facilitates the analysis of distinct learning techniques.

\subsubsection{Labels}
The label of each instance in the NSL-KDD is assigned as either normal or an attack, with exactly one specific attack type. The attacks fall in one of following categories in Table \ref{tab:attack_c} with the statistics summary of specific attack types, where \textit{R2L} represents the \textit{Remote-To-Local Attack}, \textit{U2R} represents the \textit{User-To-Root Attack}, and \textit{DoS} represents the \textit{Denial-of-Service Attack}.



\subsubsection{Features}
There are 41 features in the NSL-KDD data set that describe the characteristics of the cyber network, which can be further divided into three groups, consisting of 9 basic features, 19 traffic features, and 13 content features. The basic features involve all attributes that can be captured from a TCP/IP connection; the traffic features contain "same host" and "same service" features based on a connection window of 100 connections; the content features are related to suspicious behavior in the data portion like the number of failed login attempts. 
Generally speaking, traffic features are useful patterns to identify the DoS and probing attacks since they need to scan the hosts or send packets (many connections to some hosts) within a very short period of time. On the contrary, R2L and U2R attacks do not have any intrusion-frequent sequential patterns but are embedded in the data portions of the packets in a single connection. Hence, the content features are better patterns that can be used to detect these two attacks. A detailed explanation of each attribute is described in \cite{dhanabal2015study}. 



\subsection{Data Visualization}
To further understand and explore the NSL-KDD data set we used visualization techniques. Data visualization is the practice of graphically representing data. Using such methods we can gain insight and make better sense of large data sets By visualizing the NSL-KDD data set we gain a greater understanding of the features with relation to each other and attack type classification. For this investigation, we visualized the training data set to split up by attack type.

Our First steps in visualization are distribution histograms of the features in the data set, shown in Fig. \ref{fig:Dis}. Distribution histograms plot the value of a feature against its occurrence in the data. From these graphs, we gain valuable information on issues within the data set, redundant features, and how features relate to different attack types. One of the first things we noticed during our initial investigation was the lack of instances of U2R and R2L attack types within distribution graphs. This is due to the small number of examples of these attacks within the data set. Further, we were able to discover some redundant features within the data set. Features 20 and 21 always have a value of zero, feature 20, the number of outbound commands. With a closer investigation of individual features, we can gain some insight into how specific features correlate to specific attack types. Fig. \ref{fig:Dis} shows some examples as to how these features correlate to specific attack types. We can see that most attacks use TCP. TCP has many vulnerabilities often exploited by attackers. DOS attacks often take advantage of the TCP handshake protocol by flooding a target host with incomplete connection and service requests in the hopes to waste server resources. As the server or host is attempting to handle a large number of connections from the attacker, it is not able to handle the requests of legitimate users thus rending the host inaccessible. This is reflected in the distribution graphs of connection count, which shows a large number of connections, and Serror Rate, which shows if those connection attempts had no further replies. Another example shown in Fig. \ref{fig:Dis} is with the Dst Host Srv Diff Host Rate histogram, which shows a correlation to the probe attack type. This graph shows the percent of connections to different destination machines from the same port number. Probe attacks will provide information on what each port is doing and what is using that port by sending information and waiting for a response. This nature is reflected in this graph as Probe attacks show up in greater numbers as this feature increases.

\begin{figure}[htbp]
    \centering
    \includegraphics[scale=.33]{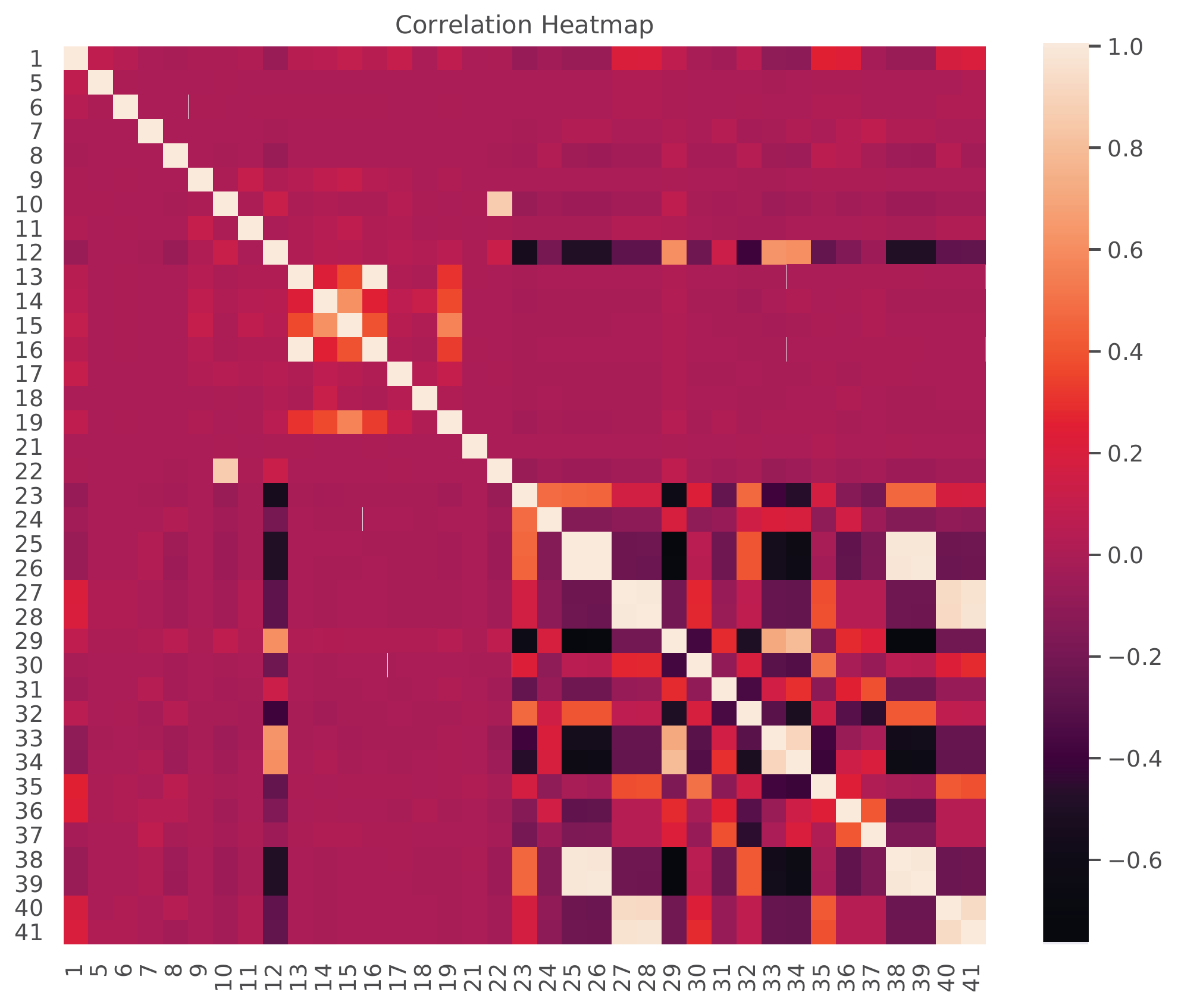}
    \caption{Correlation heat map of features in the NSL-KDD data set.}
    \label{fig:Cor}
\end{figure}

\begin{figure}[htbp]
    \centering
    \includegraphics[scale=.175]{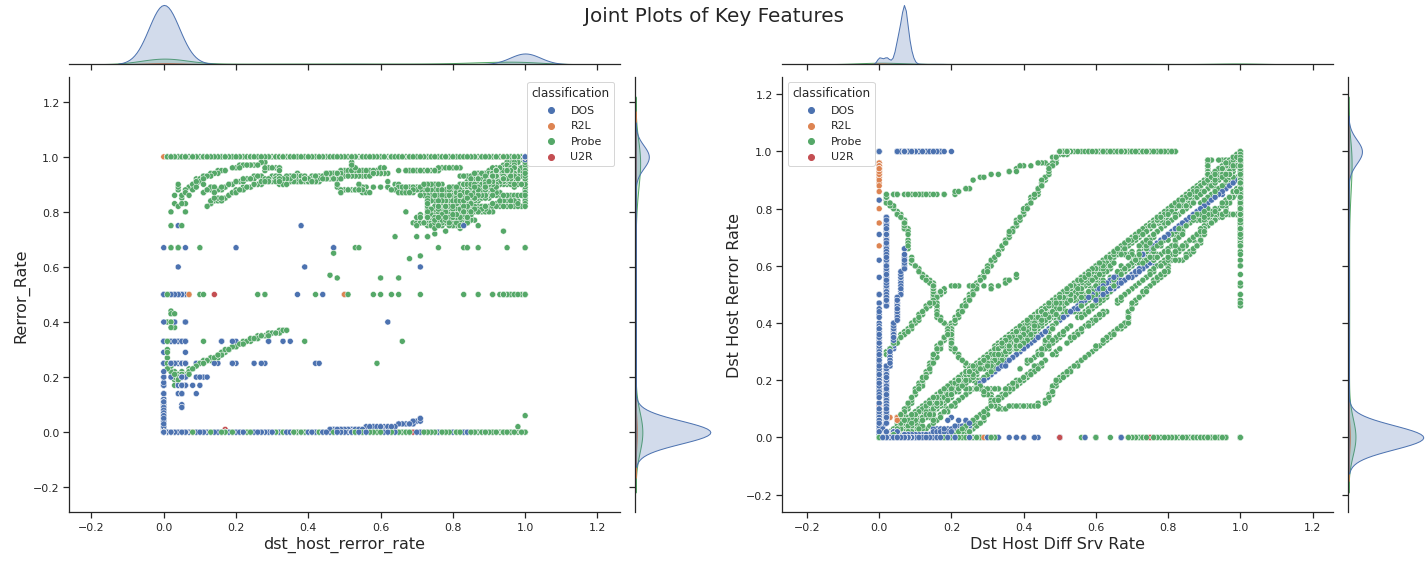}
    \caption{Scatter plots of specific features in the NSL-KDD data set.}
    \label{fig:Scat}
\end{figure}

We further investigated the data set by calculating the correlation coefficient of the features compared to each other which can be seen in Fig. \ref{fig:Cor}. This provides us insight into the strength of the relationship between the two figures. The greater the correlation the closer the value is to -1, or 1.
Fig. \ref{fig:Cor} shows a strong correlation between higher-level figures which shows that these higher-level figures have a higher potential to provide information. These higher-level figures are often based on each other which can also explain why they have such high significance.
Examples of these relationships are shown in the scatter plots of Fig. \ref{fig:Scat}. Here we can see correlation between higher level features which provide information on he probe attack type. We can see that the probe attack type often threw the REJ flag, but this percentage was often affected by the number of services the probe reached out to. The further a probe attacked reached the greater its connection attempt was rejected. By observing these relations more directly we can determine the importance of some feature pairings towards anomaly detection and attack type classification.

\section{Methodology}
In this section, we present the pipeline of our detection mechanism. Firstly, we present the preprocessing details and imbalanced nature of the explored dataset. After that, we also present a detection strategy and briefly describe the utilized machine learning algorithms. We adopt a hierarchy two-stage detection method: a binary classification followed by a 4-class classification. In the first stage, each input is classified into two normal samples and anomalies. We will explore different learning algorithms, including supervised learning classifiers and an unsupervised learning model, autoencoder. In the second stage, anomalies are further classified into four main attack categories (DoS, Probe, R2L, U2R), where supervised learning models are leveraged.

\subsection{Preprocessing}
Although the NSL-KDD data set is a cleansed data set, we still need preliminary preprocessing feature engineering before the data is fed into the model. To be specific, the categorical features should be converted into the numerical form so that they can be thought of as a vector in the Euclidean space: Three attributes (\textit{'protocol\_type','service',and 'flag'}) are categorical, we encode them by using a LabelCount encoder which sorts the categories by the frequency of each category within the feature. LabelCount has specific advantages at the outlier-insensitive nature and a reduction of dimensionality when certain features have very large numbers of categories. 

After assigning numerical values to each categorical feature, the next step is to normalize each feature, as features that are measured at different scales do not contribute equally to the analysis and can create a bias for models. Therefore, the standardization as shown in Equation \ref{eq:std} is applied to transform the data to comparable scales (around the center 0 with a standard deviation of 1).

\begin{equation}
    Z = \frac{x - \mu}{\sigma}
\label{eq:std}
\end{equation}
where $Z$ denotes the standardized feature, $x$ denotes each value within the feature, $\mu, \sigma$ denotes the	mean and standard deviation of all values, respectively.


In this dataset, the number of examples across the classes for the binary classification (normal vs. others) is roughly close. However, the 4-class intrusion data suffers from a severe imbalance, as the ratio of each class is approximately $920:220:20:1$. Most machine learning algorithms assume or expect a balanced class distribution for pattern learning and downstream classification tasks. When such data skewness exists, these algorithms fail to properly represent the distributive characteristics of the data whose results provide invalid accuracy across the classes of the data \cite{he2009learning}. To alleviate the negative effect of imbalanced data set, we employ SVM-SMOTE\cite{svmsmote}, a sophisticated oversampling technique leveraging a Support Vector Machine algorithm to detect sample to use for generating new synthetic samples under the framework of Synthetic Minority Oversampling Technique (SMOTE) class \cite{chawla2002smote}.  

\begin{figure}[htbp]
    \centering
    \includegraphics[scale=.53]{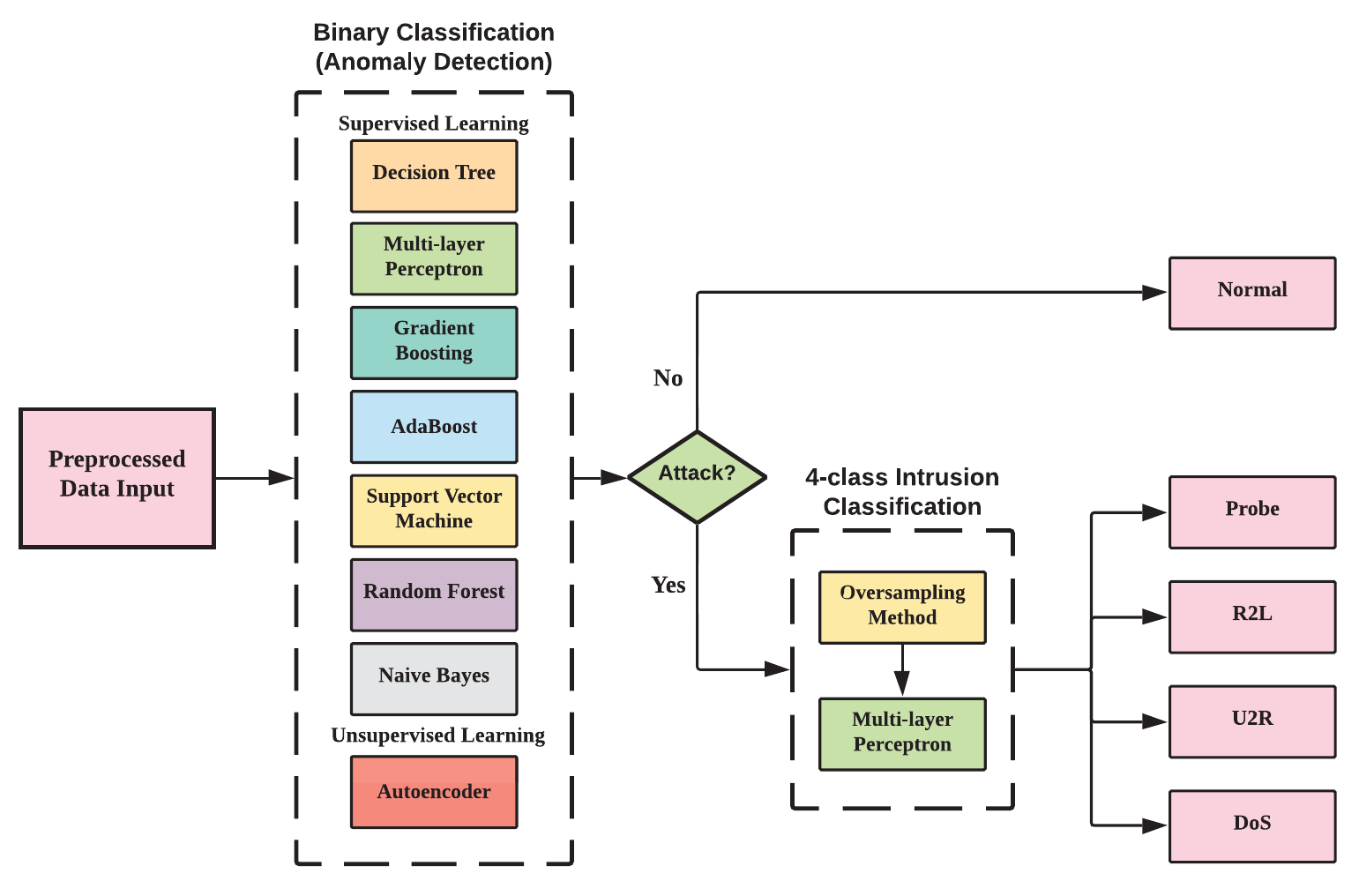}
    \caption{Learning-based on hierarchy intrusion detection strategy.}
    \label{fig:overview}
\end{figure}

\subsection{Learning to Detect Network Intrusion}

In this paper, we adopt various learning models for binary and 4-class intrusion detection. Fig. \ref{fig:overview} depicts the overview of our IDS based on a hierarchy machine learning strategy. For binary detection, we utilize supervised learning models including Decision Tree, Random Forest, Naive Bayes, Support Vector Machine (SVM), AdaBoost, Gradient Boosting, Multi-layer Perceptron (MLP). Besides, we also consider an unsupervised representation learning model, Autoencoder, and treat the conventional binary classification problem as an anomaly detection problem. It learns the representation of normal samples and uses the reconstruction error as the anomaly score. At the testing stage, samples with high reconstruction (exceeding the threshold) are considered anomalies, as it is assumed that anomalies are difficult to be reconstructed \cite{an2015variational}. The specific algorithm of Autoencoder-based detection is shown in Algorithm 1. After anomalies are detected, we utilize a deep neural network accompanied with the aforementioned oversampling technique to classify the specific intrusion type in a more robust manner.

 \begin{algorithm}[H]
 \caption{Autoencoder-based attack detection algorithm}
 \begin{algorithmic}[1]
 \renewcommand{\algorithmicrequire}{\textbf{Input:}}
 \renewcommand{\algorithmicrequire}{\textbf{Parameters:}}
 \renewcommand{\algorithmicensure}{\textbf{Output:}}
 \REQUIRE Normal dataset $\mathbf{X}$, Anomalous dataset $x^i$, $i = 1, ... N$, threshold $\alpha$ defined by validation loss
 \REQUIRE $f_{\theta}$ : Encoder, $g_{\phi}$ : Decoder
 \ENSURE  \text { reconstruction errors, anomaly indicator } 
  \STATE $g_{\phi}, f_{\theta} \leftarrow$ train a Autoencoder with normal dataset $\mathbf{X}$.
 \\ \textit{LOOP Process}
  \FOR {$i = 1$ to $N$}
  \STATE $\text { reconstruction error }(i)=\left\|x^{(i)}-g_{\theta}\left(f_{\phi}\left(x^{(i)}\right)\right)\right\|^2$
  \IF {reconstruction error $> \alpha$} 
    \STATE $x^i$ is an anomaly (attack)
  \ELSIF {reconstruction error $<= \alpha$} 
    \STATE $x^i$ is not an anomaly (attack)
  \ENDIF
  \ENDFOR
 \label{alg:autoencoder}
 \end{algorithmic}
 \end{algorithm}

\section{Experiments}
In this section, we evaluate the performance of different machine learning algorithms for two hierarchic stages of our intrusion detection system. We also explore the effectiveness of SVM-SMOTE oversampling technique toward a more valid classification model.

\subsection{Evaluation Metrics and Experiment Settings}

In this paper, we adopt accuracy, precision, recall, and F1 score for a comprehensive evaluation in the binary classification task. For 4-class classification, we use accuracy, F1 score of each intrusion type (one vs. all), with their macro-average (arithmetic mean) and micro-average (weighted mean) to demonstrate the classification performance of our model with the existence of imbalance \cite{liu2017efficient}.   
\begin{equation}
    \text{Accuracy} = \frac{\mathrm{Number\; of \;correct\; predictions}}{\mathrm{Total\; number\; of\; predictions}}
\end{equation}
\begin{equation}
\text{Precision} =\frac{\mathrm{TP}}{\mathrm{TP}+\mathrm{FP}}
\end{equation}
\begin{equation}
\text{Recall} =\frac{\mathrm{TP}}{\mathrm{TP}+\mathrm{FN}}
\end{equation}
\begin{equation}
\mathrm{F1} = 2 \times \frac{\text { Precision } \times \text { Recall }}{\text { Precision }+\text { Recall }}
\end{equation}
where $\mathrm{TP}$ = True Positives, $\mathrm{TN}$ = True Negatives, $\mathrm{FP}$ = False Positives, and $\mathrm{FN}$ = False Negatives.

Next, we present the settings of the experiment for Autoencoder in binary classification and deep neural network in multi-class classification. The Autoencoder has three layers with 15 neurons in hidden space and 0.15 Gaussian Noise, and 0.05 Dropout rate in encoding layers for regularization. The activation for both encoder and decoder is Scaled Exponential Linear Unit (SeLU), and the loss function is Mean Squared Error (MSE). In a deep neural network, three layers are used, where there are 80 neurons in the hidden layer with Rectified Linear Unit (ReLU) as activation function, four neurons in the output layer with Softmax as activation function. The loss function is cross-entropy. For both tasks, the batch size is 32; 0.15 of training data are used for validation; the optimizer is Adam with a learning rate 0.001; early stopping is adopted with the patience of 6 steps.

\subsection{Binary Classification}

\begin{table} 
  \setlength{\tabcolsep}{4.5pt}
  \small
  \caption{Performance Evaluation of Binary Classification in IDS}
  \label{Tlb: 2}
  \centering
 
     \begin{tabular}{lcccccl}
     \toprule
     \cmidrule{1-5}
     {Model} & {Accuracy} & {Precision} & {Recall} & {F1 Score}\\

    \midrule
     Decision Tree & 68.28\%  & 68.16\%    & 83.09\%  & 0.7489  \\
     Random Forest& 76.00\% & 87.34\%   & 67.65\%  & 0.7624  \\
     Naive Bayes & 76.86\%  & \underline{96.21\%}  & 59.95\% & 0.7387 \\
     SVM & \underline{80.47\%}  & \textbf{97.56\%}  & 67.38\% & 0.7971 \\
     AdaBoost  & 79.40\%  & 86.90\%  & 75.14\% & \underline{0.8059} \\
     Gradient Boosting & 68.12\%  & 65.04\%  & \textbf{95.13}\% & 0.7726 \\
     MLP & 77.90\%  & 95.82\%  & 63.96\% & 0.7671 \\
     \midrule
     \textbf{Autoencoder} & \textbf{87.52\%}  & 93.20\%  & \underline{84.22\%} & \textbf{0.8848} \\
    \bottomrule
    \cmidrule{1-5}
    \end{tabular}

\end{table}

The evaluation of different models for binary classification is shown in Table \ref{Tlb: 2}, where the best result for each metric is indicated in bold while the corresponding second-best result is underlined. Among these learning models, it can be observed that SVM yields the highest precision score and the second highest accuracy, while Gradient Boosting classifier demonstrates its advantage on the highest recall score with a clear margin to the second highest one. In terms of the F1 score, AdaBoost and SVM achieve similarly good performance among supervised learning models. The above results illustrate that SVM and boosting methods gain more favors at separating attacks from normal samples within the scope of supervised learning.

\begin{figure}
    \centering
    \includegraphics[scale=.28]{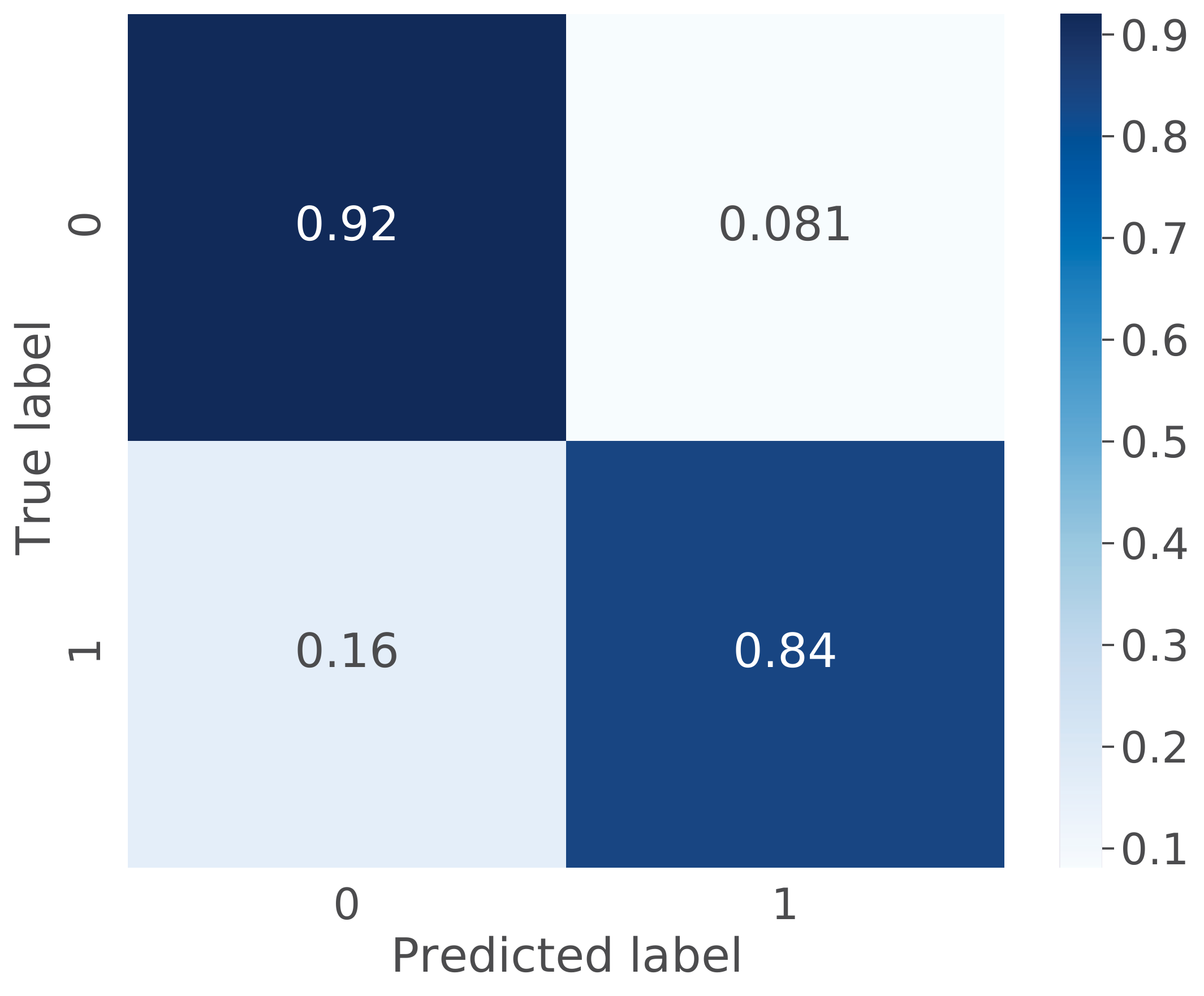}
    \caption{Confusion matrix of Autoencoder based binary classifier.}
    \label{fig:cm_binary}
\end{figure}

On the other hand, the Autoencoder shows a huge advantage in the binary classification task, as it yields the highest accuracy, F1 score, and the second-highest recall score and. Moreover, the increase of accuracy and F1 score from the second-best result is 7.05\% and 0.0789, respectively, demonstrating the power of unsupervised representation learning. To evaluate the classification performance with more details, the confusion matrix of Autoencoder is also presented, as shown in Fig.\ref{fig:cm_binary}. It is clear that Autoencoder performs slightly better on identifying normal behaviors than malicious attacks. 

In general, the implemented Autoencoder with Gaussian noise and dropout as the regularization method is a good alternative in binary classification and anomaly detection. It is able to extract the feature information and generate salient and generalized vector representations for the reconstruction of normal samples only, where the anomalies usually do not share the similar representation space and fail to be reconstructed with a much higher loss during the model inference.

\subsection{4-Class Intrusion Type Classification}

At this stage, the samples identified as intrusion are further classified into different types. Table \ref{tab:4class} shows the performance evaluation using a deep neural network for 4-class classification, without or with the aforementioned SVM-SMOTE oversampling method. It can be observed that the oversampling method has a minor impact on accuracy and micro F1 score, with the value around 80.4\% and 0.783, respectively. However, it is clear that the new synthetic samples generated by SVM-SMOTE help the model to learn patterns and significantly improves the F1 score of U2R and yields an increase of macro F1 score by 0.1, with a slightly better F1 score of R2L and little inferior scores for DoS and Probe.

\begin{figure}
    \centering
    \includegraphics[scale=.28]{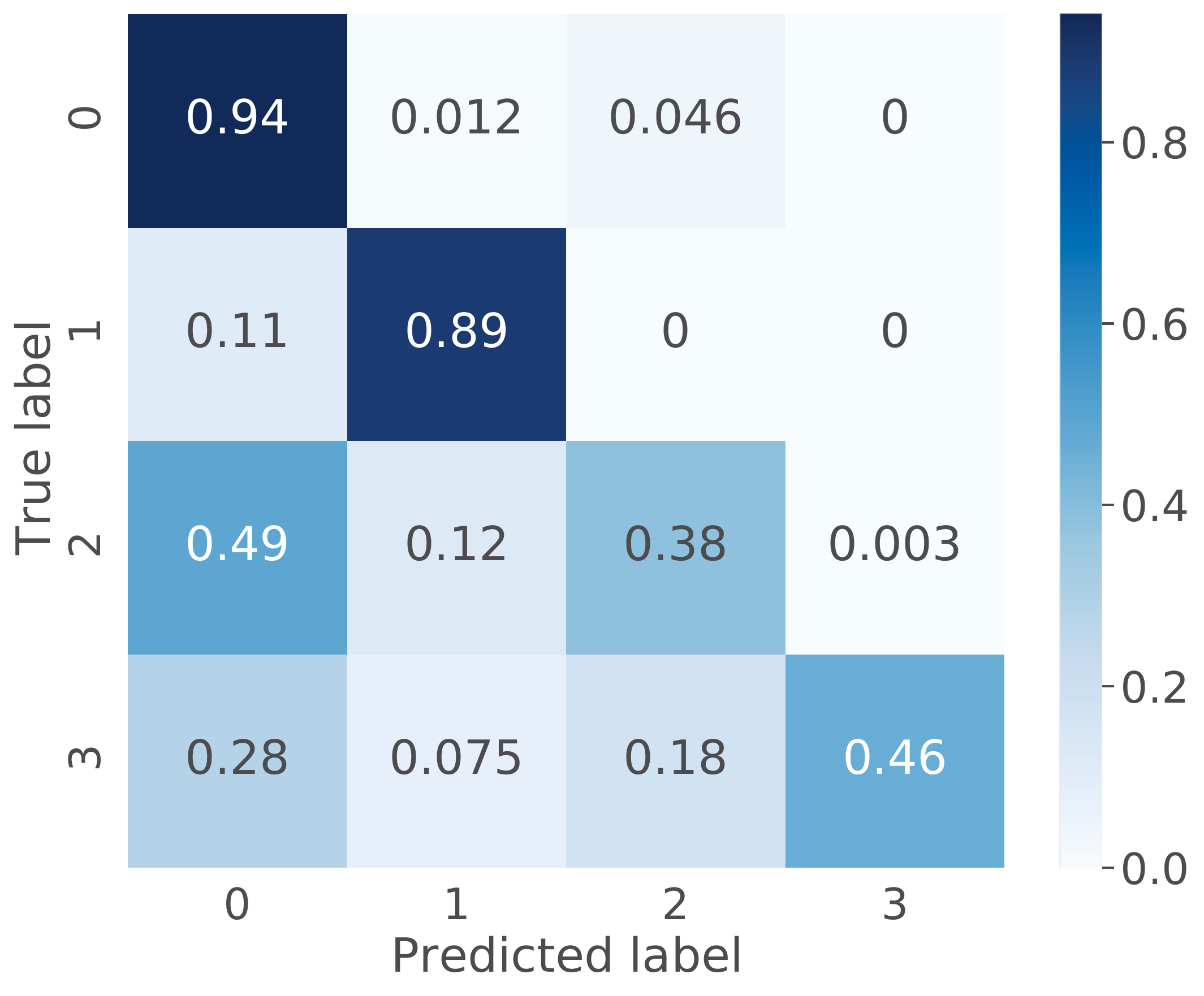}
    \caption{Confusion matrix of deep neural network model for 4-class classification.}
    \label{fig:4class}
\end{figure}

From Fig. \ref{fig:4class}, the confusion matrix of a 4-class classifier, we can conclude that despite the models are trained using the balanced data oversampled by SVM-SMOTE, it can only provide relatively accurate prediction on DoS (label 0) and Probe attack (label 1). In addition, a large portion of misclassified samples in R2L (label 2) and U2R (label 3) fall in the DoS attack, which suggests the inferior of our model at identifying the different patterns between DoS and R2L/U2R. Besides, a certain amount of misclassified samples in U2R fall in R2L, indicating a similar problem. One of the possible explanations is that, even with SVM-SMOTE to alleviate the imbalance problem for R2L and U2R, the patterns of these new synthetic samples appear to be insufficient to represent the attack behaviors in the testing set.

\begin{table}
\caption{Performance Evaluation of 4-class Classification in IDS}
\begin{center}
\begin{tabular}{  lcccl }
 \toprule
 \textbf{Evaluation Metrics} &  \textbf{w/o Oversampling} & \textbf{w/ Oversampling}\\
 \midrule
 \textbf{Accuracy} & 80.47\% & 80.36\% \\ 
 \midrule
 F$1_{DoS}$ & 0.8790 & 0.8678\\

 F$1_{Probe}$ & 0.8703 &0.8560 \\
 
 F$1_{R2L}$ & 0.4789 & 0.5100\\
  
 F$1_{U2R}$ & 0.2075 & 0.5741\\
  \midrule
 \textbf{Macro F1} & 0.6089 & 0.7020 \\
 \textbf{Micro F1} & 0.7839 & 0.7836\\
 \bottomrule
\end{tabular}
\label{tab:4class}
\end{center}
\end{table}

\section{Conclusion and Future Work}
In this paper, we proposed a data-driven intrusion detection framework based on data analytics and hierarchical learning-based detecting strategies.
Firstly, we visualize potential patterns and discuss the relationships between features and underlying attack behaviors based on domain knowledge.  
Then, we leverage the supervised and unsupervised learning method to classify normal network behaviors and malicious attacks, and demonstrate the advantage of the unsupervised representation learning model in our task. Next, we focus on the problem of imbalance and alleviate it with SVM-SMOTE oversampling technique. We further demonstrate the effectiveness and the drawback of the oversampling mechanism and in 4-class classification with a deep neural network as a base model. In general, our framework yields satisfactory results on classifying normal samples and samples of the DoS and Probe attacks. However, it still shows certain inferiority in terms of minority class with an oversampling technique. For future work, we consider using more sophisticated learning models and techniques in terms of data imbalance, for example, cost-sensitive learning. 

\section*{Acknowledgment}
This research was supported by the National Science Foundation under Grant No. 1956193.

\bibliographystyle{./bibliography/IEEEtran}
\bibliography{./bibliography/IEEEabrv,./bibliography/IEEEexample}


\end{document}